\documentclass[fleqn,usenatbib]{mnras}

\usepackage[T1]{fontenc}
\DeclareRobustCommand{\VAN}[3]{#2}
\let\VANthebibliography\thebibliography
\def\thebibliography{\DeclareRobustCommand{\VAN}[3]{##3}\VANthebibliography}
\usepackage{graphicx}
\usepackage{amsmath, amssymb}
\usepackage{threeparttable,multirow}
\usepackage{newtxtext,newtxmath}
\usepackage{color,soul}

\def\kms{\,km\,s$^{-1}$}

\def\re{$R_{\rm e}$}
\def\mstar{$\rm M_{*}/M_{\odot}$}
\def\sigs{$\rm \sigma_{S}$}
\def\sigb{$\rm \sigma_{B}$}
\def\sign{$\rm \sigma_{N}$}
\def\fb{$f_{\rm B}$}

\def\cagn{$\rm C_{AGN}$}
\def\cagns{$\rm C_{AGN,S}$}
\def\cagnn{$\rm C_{AGN,N}$}
\def\cagnb{$\rm C_{AGN,B}$}

\def\sfden{$\rm \Sigma_{SFR}$}
\def\sfdens{$\rm \Sigma_{SFR,S}$}
\def\sfdenn{$\rm \Sigma_{SFR,N}$}
\def\sfdenb{$\rm \Sigma_{SFR,B}$}

\title[Feedback and gas velocity dispersions]{The SAMI Galaxy Survey: impact of star formation and AGN feedback processes on the ionized gas velocity dispersion}

\author[S.~Oh et~al.]{Sree~Oh,$^{1,2,3}$\thanks{E-mail: sree.oh@yonsei.ac.kr} 
Matthew~Colless,$^{2,3}$
Stefania~Barsanti,$^{2,3}$
Henry R. M. Zovaro,$^{2,3}$
Scott M. Croom,$^{4,3}$
\newauthor
Sukyoung K. Yi,$^{1}$
Andrei Ristea,$^{5, 3}$
Jesse van de Sande,$^{6,4,3}$
Francesco D'Eugenio,$^{7}$
Joss, Bland-Hawthorn,$^{4,3}$
\newauthor
Julia J. Bryant,$^{4,3}$
Sarah Casura,$^8$
Hyunjin Jeong,$^9$
Sarah M. Sweet,$^{10,3}$
Tayyaba Zafar $^{11,12}$
\\ ~ \\
$^{1}$Department of Astronomy and Yonsei University Observatory, Yonsei University, Seoul 03722, Republic of Korea\\
$^{2}$Research School of Astronomy and Astrophysics, Australian National University, Canberra, ACT 2611, Australia\\
$^{3}$ARC Centre of Excellence for All Sky Astrophysics in 3 Dimensions (ASTRO~3D), Australia\\
$^{4}$Sydney Institute for Astronomy (SIfA), School of Physics, The University of Sydney, NSW 2006, Australia\\
$^{5}$International Centre for Radio Astronomy Research, University of Western Australia, 35 Stirling Highway, Crawley, WA 6009\\
$^{6}$School of Physics, University of New South Wales, Sydney, NSW 2052, Australia\\
$^{7}$Cavendish Laboratory and Kavli Institute for Cosmology, University of Cambridge, Madingley Rise, Cambridge, CB3 0HA, United Kingdom\\
$^{8}$Hamburger Sternwarte, Universitaet Hamburg, Gojenbergsweg 112, D-21029 Hamburg, Germany\\
$^{9}$Korea Astronomy and Space Science Institute, Daejeon 34055, Republic of Korea\\
$^{10}$School of Mathematics and Physics, University of Queensland, Brisbane, QLD 4072, Australia\\
$^{11}$School of Mathematical and Physical Sciences, Macquarie University, NSW 2109, Australia\\
$^{12}$Macquarie University Astrophysics and Space Technologies Research Centre, Sydney, NSW 2109, Australia
}

\date{Accepted XXX. Received YYY; in original form ZZZ}

\begin{document}
\label{firstpage}
\pagerange{\pageref{firstpage}--\pageref{lastpage}}\pubyear{2021}
\maketitle

\begin{abstract}
We investigate the influence of star formation and instantaneous AGN feedback processes on the ionized gas velocity dispersion in a sample of 1285 emission-line galaxies with stellar masses $\log\,(M_*/M_{\odot}) \geq 9$ from the integral-field spectroscopy SAMI Galaxy Survey. We fit both narrow and broad emission line components using aperture spectra integrated within one effective radius, while ensuring the elimination of velocity differences between the spectra of individual spaxels. Our analysis reveals that 386 (30\%) galaxies can be adequately described using a single emission component while 356 (28\%) galaxies require two (broad and narrow) components. Galaxies characterized by high mass, elevated star formation rate surface density, or type-2 AGN-like emissions tend to feature an additional broad emission-line component, leading to their classification as double-component galaxies. We explore the correlations between $M_*$ and gas velocity dispersions, highlighting that the prominence of the broad component significantly contributes to elevating the gas velocity dispersion. Galaxies displaying AGN-like emission based on optical definitions show enhanced gas velocity dispersions. In star-forming galaxies, both stellar mass and star-formation rate surface density substantially contribute to the velocity dispersion of the narrow component. Increased star-forming activity appears to elevate the velocity dispersion of the narrow component. The broad component exhibits a weaker dependence on stellar mass and is primarily driven by galactic outflows. We suggest that strong star forming activity leads to the formation of a broad emission-line component, but the impact on inflating gas velocity dispersion is moderate. On the other hand, AGN-driven outflows appear to be a more important contributor to the elevated velocity dispersion of the ionized gas.

\end{abstract}


\section{Introduction}
\label{sec:intro}
Galactic winds are powerful gas streams expelled by the momentum produced in various feedback processes, including starbursts and active galactic nuclei (AGN). The association between these feedback processes and galactic winds has been firmly established, supported by the observation of high-velocity outflows and the extended distribution of metal-enriched gas within galaxies (e.g.\ Veilleux et~al.\ 2005; Bland-Hawthorn et~al.\ 2007; Dalcanton 2007; Weiner et~al., 2009; Fabian et~al.~2012; Cicone et~al.\ 2014; Chisholm et~al.\ 2018; Rupke 2018; Ginolfi 2020; Woo, Son, \& Rakshit 2020). Furthermore, theoretical models and simulations have consistently provided predictions that align with these observations (e.g.\ Cooper et~al.~2008; Dalla Vecchia et~al.\ 2008, 2012; Wiersma et~al.\ 2009; Muratov et~al.\ 2015; Tescari et~al.\ 2018; Zhang 2018; Nelson et~al.\ 2019). Feedback-induced processes in the form of galactic winds play several important roles in galaxy evolution, including the regulation of star formation (e.g.\ Hayward \& Hopkins 2017), the redistribution of gas and metals (e.g.\ Choi et~al.\ 2020), and the enrichment of the surrounding intergalactic medium with heavy elements (e.g.\ Shen, Wadsley, \& Stinson 2010). 

Galaxies that are associated with galactic winds often display gas emissions characterized by multiple components (e.g.\ McElroy et~al.\ 2015; F{\"o}rster Schreiber et~al.\ 2019; Freeman et~al.\ 2019; Rodr{\'\i}guez del Pino et~al.\ 2019; Wylezalek et~al.\ 2020; Couto et~al.\ 2021; Fu et~al.\ 2023; Llerena et~al.\ 2023; Kim et~al. 2023). These components are typically distinguished based on the width of their line profile (e.g.\ narrow or broad). Broad emission-line components frequently serve as indicators of galactic winds, providing evidence of feedback processes. Recently, galactic wind studies have been facilitated by the use of spatially-resolved spectra obtained through integral-field spectroscopy (IFS). Ho et~al.\ (2014) performed spectral decomposition on a starburst galaxy included in the Sydney-AAO Multi-object Integral-field (SAMI; Croom et~al.\ 2012) survey. Their analysis revealed multiple emission-line components associated with HII regions and shock excitations. Zovaro et~al.\ (2014) extended the spectral decomposition to a large IFS dataset of star-forming galaxies, indicating that galaxies characterized by high star-formation rate surface density tend to exhibit multiple spectral components. In the study by Avery et~al.\ (2021), aperture spectra were decomposed into broad and narrow line components using data from the Mapping Nearby Galaxies at Apache Point Observatory (MaNGA; Bundy et~al.\ 2015) IFS survey. They statistically confirmed that both star formation and AGN activities correlate with the gas outflow rates in 322 candidates for gas outflows, particularly when derived using the broad emission-line component.

Ionized gas shows distinct kinematic properties depending on the ionization sources (Ho et~al.\ 2014; Rich, Kewley, \& Dopita 2015; Woo, Son \& Bae 2017). In a recent study, Oh et~al.\ (2022) provided statistical confirmation that the velocity dispersion of ionized gas, when normalized by the stellar velocity dispersion, correlates with the contribution of AGN-like emission determined from optical emission-line diagnostics (e.g.\ Baldwin, Phillips \& Terlevich 1981; Kewley et~al.\ 2001; Kauffmann et~al.\ 2003; Kewley et~al.\ 2006). This result suggests that the kinematics of ionized gas (in contrast to stars) are intimately connected to the dominant source of ionization, aligning with the predictions made by Belfiore et~al.\ (2016) and supported by Law et~al.\ (2021). Different kinematic signatures observed in the narrow and broad emission line components may explain a close correspondence between excitation sources and their integrated kinematics, further confirming the influence of feedback processes on gas motions. 

The goal of this study is to investigate the influence of active star formation and AGN on the ionized gas kinematics in various types of galaxies. We analyze the velocity dispersion of narrow and broad emission line components and elucidate the factors influencing their kinematics, taking advantage of spatially and spectrally resolved emission line profiles from the SAMI IFS data. It is essential to comprehend the intrinsic velocity dispersion of the narrow line component across a large sample of galaxies and confirm whether the narrow line components from various types of galaxies share common characteristics. Recent IFS studies have reported positive correlations between stellar mass and gas velocity dispersion (Cortese et al. 2014; Barat et al. 2019; Rodríguez del Pino et al. 2019; Varidel et al. 2020), suggesting that gas kinematics may still be partially linked to gravitational potential, despite their susceptibility to feedback processes. Therefore, stellar mass--kinematic relations serve as a useful tool to assess the homogeneity of the narrow-line component and to determine whether it is predominantly described by gravitational potential or also affected by feedback processes. Another open question is whether the velocity dispersion of the narrow line component change based on the dominant ionization sources within a galaxy. Once the properties of the narrow line component are established, the subsequent step is to examine the presence and contribution of broad line components to elucidate the impact of feedback processes on the velocity dispersion of ionized gas.

The paper is organized as follows. We provide an overview of the SAMI Galaxy Survey in Section 2. The establishment of velocity-corrected spectra and decomposition of narrow and broad line components are outlined in Section 3. Emission line diagnostics and mass-kinematic relations are presented in Section 4. In Section 5, we discuss the factors influencing the kinematics of the narrow and broad line components, and our conclusions are summarized in Section 6. Throughout the paper, we adopt a standard $\Lambda$CDM cosmology with $\Omega_{\rm m} = 0.3$, $\Omega_{\Lambda} = 0.7$, and $H_0 = 70$\kms~Mpc$^{-1}$.

\section{The SAMI Galaxy Survey}
\label{sec:sami}
SAMI is a multi-object fibre integral field system feeding the AAOmega dual-arm spectrograph on the Anglo-Australian Telescope (Sharp et~al.\ 2006). SAMI employs 13 hexabundles, each composed of 61 optical fibres, each 1.6~arcsec in diameter (Bland-Hawthorn et~al.\ 2011; Bryant et~al.\ 2011, 2014). Blue (3750--5750\,\AA) and red (6300--7400\,\AA) arms use 580V and 1000R gratings, which return spectral resolutions $R$=1808 and $R$=4304, respectively (van de Sande et~al.\ 2017).

The third and final data release (DR3, Croom et~al.\ 2021) of the SAMI Galaxy Survey includes 3068 unique galaxies at redshifts $0.04 < z < 0.095$ (Bryant et~al.\ 2015), from the three equatorial regions of the Galaxy And Mass Assembly (GAMA; Driver et~al.\ 2011) survey and additional eight galaxy clusters (Owers et~al.\ 2017). We use the data cubes and the measurements of ionized gas and stellar kinematics from SAMI DR3; see van de Sande et~al.\ (2017) for details on the measurement of stellar kinematics. The stellar masses ($\rm M_*/M_\odot$) have been determined using $i$-band magnitudes and $g-i$ colours, as described in the works of Taylor et~al.\ (2011) and Bryant et~al.\ (2015). The effective radii \re\ of the SAMI galaxies have been measured applying the Multi Gaussian Expansion algorithm (MGE; Emsellem et~al.\ 1994) to the Sloan Digital Sky Survey (SDSS; York et~al.\ 2000) and the VLT Survey Telescope (VST) ATLAS Survey (Shanks et~al.\ 2013) imaging data; see D'Eugenio et~al.\ (2021) for more information on the MGE measurements of the SAMI galaxies. 

\section{Methods}
We perform fittings for both narrow and broad emission line components in SAMI aperture spectra. The subsequent subsections describe in detail a series of steps for eliminating velocity differences attributed to gas rotations, generating aperture spectra, subtracting the continuum, employing {\sc lzifu} for fitting emission lines, identifying galaxies hosting a reliable broad emission line component, and finalizing the sample selection.

\subsection{Velocity-field corrected aperture spectra}
\label{sec:spectra}
We generate aperture spectra integrated within an elliptical 1 \re\ aperture using the data cubes from the SAMI DR3 for a reliable fit of narrow and broad line components. However, simple integration inevitably broadens emission lines by summing spectra with different radial velocities. Moreover, aperture spectra sometimes show strong substructures in their emission lines according to the distribution of the radial velocity fields within the aperture. The velocity structures embedded in the integrated emission lines obstruct an accurate estimate of narrow and broad line components. We therefore eliminate the velocity differences between spaxels before we integrate the spectra to prevent line broadening caused by velocity fields while maintaining a high S/N (Rosales-Ortega et~al.\ 2012; Swinbank et~al.\ 2019; Avery et~al.\ 2021).

\begin{figure}
\centering
\includegraphics[width=\columnwidth]{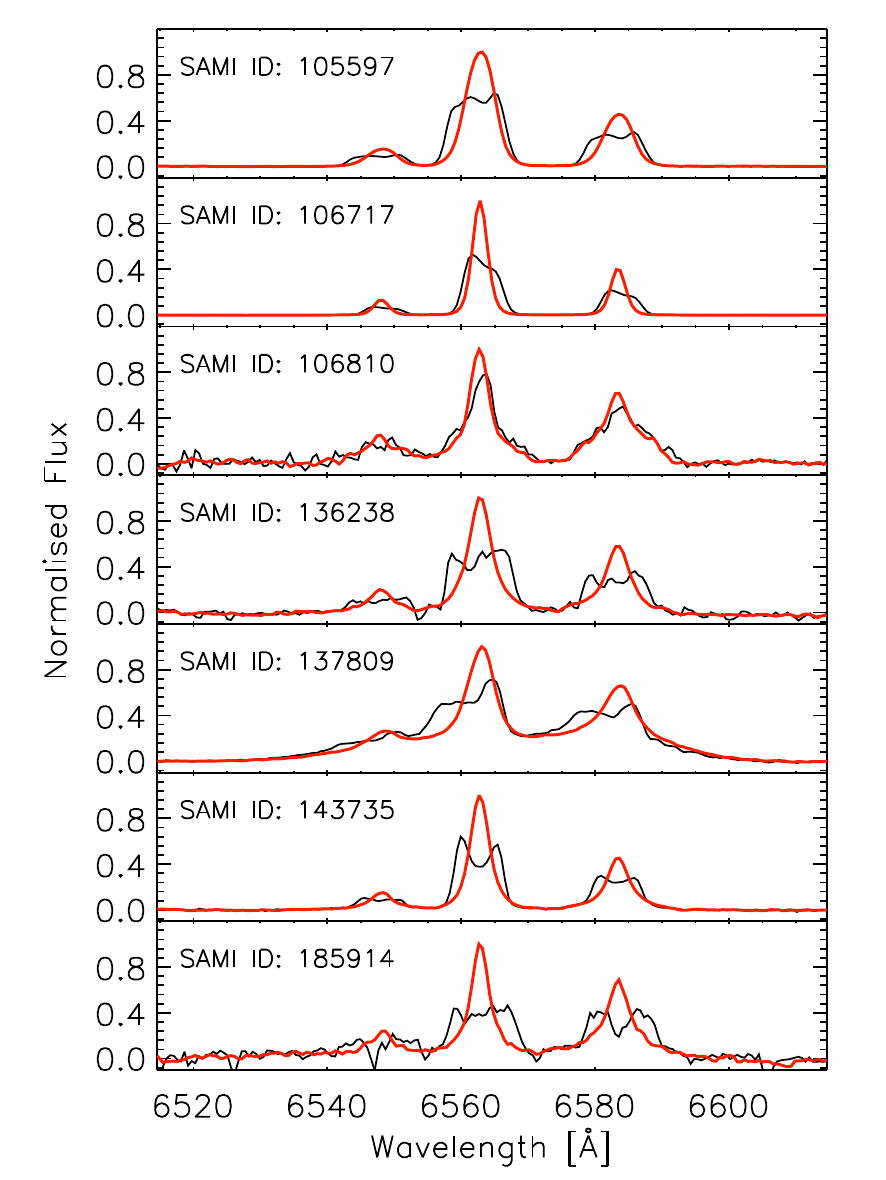}
\caption{Continuum-subtracted aperture spectra within 1\re\ for 7 SAMI galaxies. The emission lines display complex structures when integrating spectra without eliminating the velocity differences between individual spectra (black line). However, when the difference in the radial velocities are removed before integration (red line), these structures are no longer apparent and the velocity-corrected aperture spectra exhibit significantly improved amplitude-to-noise ratio in the emission lines.}
\label{aper}
\end{figure}

We used ionized gas rotation velocities published in the SAMI DR3 as a reference point to eliminate the relative velocity differences between individual spectra. The SAMI team measured the rotation velocity of ionized gas for individual spaxels using the emission line fitting code {\sc lzifu} (Ho et~al.\ 2016), after subtracting the continuum that is estimated using the Penalized Pixel-Fitting method ({\sc pPXF}; Cappellari \& Emsellem 2004; Cappellari 2017). Owers et~al.\ (2019) provide more details on the measurement of the continuum. We shifted spectra from each spaxel to remove the relative differences in their rotation velocities and then integrated the shifted spectra within 1~\re. In Figure~\ref{aper}, we present example H$\alpha$ and [NII] emission lines to show the difference between aperture spectra with and without aligning the velocity fields. Integrated emission lines without removing rotation show complex structures that are difficult to describe as sums of narrow and broad line components. Moreover, there is a high chance of artificially introducing a broad component with such complex emission line structures. In contrast, emission lines integrated after removing the velocity differences do not show such complex structures.

\begin{figure}
\centering
\includegraphics[width=\columnwidth]{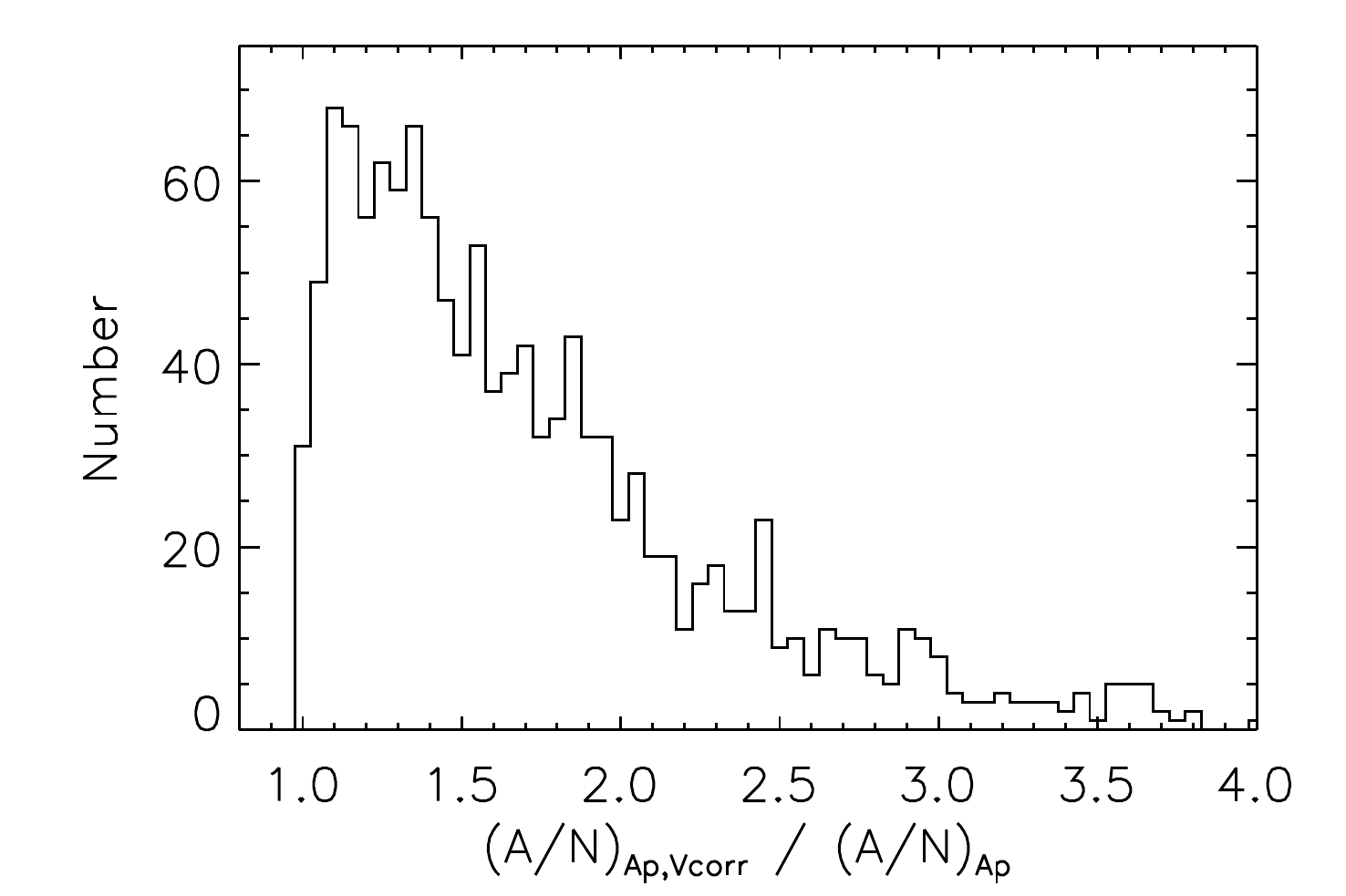}
\caption{The comparison of the amplitude-to-noise ratios, A/N, of H$\alpha$ measured from the velocity-corrected aperture spectra, $\rm (A/N)_{Ap,Vcorr}$, and the aperture spectra without the velocity correction, $\rm (A/N)_{Ap}$.}
\label{an}
\end{figure}

We use the peak amplitude-to-noise (A/N) ratio of the H$\alpha$ emission line flux to assess the influence of the velocity correction on the emission line within the aperture spectra, given that the S/N derived from the continuum is less informative about the strength of emission lines. We note that the noise was locally estimated over the wavelength range of 6300--6900\,\AA. In Figure~\ref{an}, we present the comparison of the H$\alpha$ A/N between two aperture spectra: one with the velocity correction applied and the other without the correction. The aperture spectra that underwent the velocity correction consistently show an improved H$\alpha$ A/N compared to the ones without the correction. The clean distribution and the higher A/N of emission lines from the velocity-corrected aperture spectra justify correcting the velocity differences between individual spectra within 1\re.

\subsection{Emission line fitting}
\label{sec:fit}

We generated velocity-field-corrected aperture spectra for 2955 galaxies, excluding nine galaxies from the initial 2964 primary target sample of the SAMI survey due to the absence of MGE \re\ measurements. We applied a stellar mass cut of $10^{9}\,{\rm M_\odot}$ for a reliable measure of gas kinematics, given the spectral resolution limit of SAMI ($\sim$29.9 km/s in red arm; van de Sande et~al.\ 2017) and low signal-to-noise (S/N) in the low mass galaxies, resulting in a sample size of 2517. We then select 1285 emission galaxies whose H$\alpha$, H$\beta$, [N\,{\small II}], and [O\,{\small III}] A/N ratios are greater than 5.

\begin{figure}
\centering
\centering\includegraphics[width=\columnwidth]{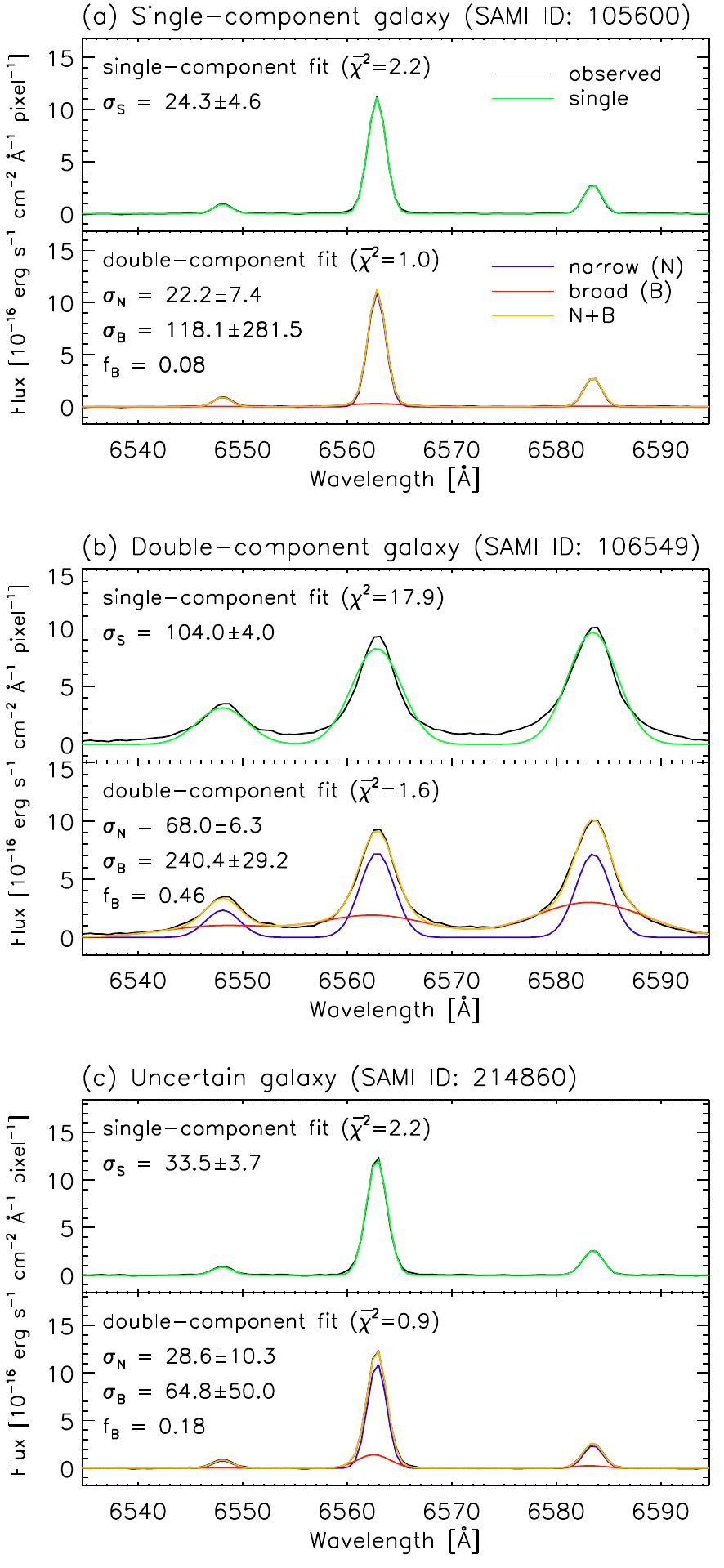}
\caption{Example of emission line fitting using {\sc lzifu} with single and double Gaussian components for (a) single-component, (b) double-component, and (c) uncertain galaxy. The upper panels show the continuum-subtracted spectra (black) and the emission-line model with a single-component fit (green) near the H$\alpha$ $\lambda$6563\AA~and [N{\small II}] $\lambda$6583\AA~emission lines. In the lower panels, model spectra for narrow (blue) and broad (red) components, along with the reconstructed total spectrum model (yellow) obtained from the double-component fit, are presented.}
\label{lzifu}
\end{figure}

We utilized the {\sc pPXF} method with the MILES simple stellar population (SSP) library (Vazdekis et al. 2010) and younger SSP templates from Gonz\'alez Delgado et~al. (2005) to estimate the continuum and subtracted it from the integrated spectra, as described in Owers et~al. (2019). Subsequently, we employed {\sc lzifu} to perform a simultaneous fit of strong emission lines, including [O{\small II}] $\lambda$3727+3729, H$\beta$, [O{\small III}] $\lambda$5007, H$\alpha$, [N{\small II}] $\lambda$6583, [S{\small II}] $\lambda$6716, and [S{\small II}] $\lambda$6731. We performed spectral fitting twice using {\sc lzifu}, employing both single and double Gaussian components. Zovaro et al. (2024) introduced a third extra-broad emission component to the emission line fitting and reported that galaxies with such components are rare. Therefore, we decided not to include the third components in our analysis. The mean velocities and velocity dispersions were constrained to be the same for all the emission lines, while the relative strengths of the emission lines were allowed to vary. Additionally, we computed the reduced $\chi^2$ values for both the single- and double-component fits ($\rm\overline\chi^2_{S}$ and $\rm\overline\chi^2_{D}$) to assess the goodness of fit. Figure~\ref{lzifu} presents example of the emission line fitting performed using {\sc lzifu} with single and double Gaussian components. 

When conducting the double component fitting, we assigned the label `narrow' to the results from the component exhibiting the lower velocity dispersion, while the other component was designated the `broad' component. We employed several initial guesses for the kinematics of the narrow and broad models. We assign the initial velocity for the narrow component as the systemic velocity derived from the redshift listed in the SAMI input catalogue (Bryant et al. 2015; Owers et al. 2017). Following this, we specify initial velocities for the broad component to be -50, 0, and +50 km/s relative to the initial velocity of the narrow component. We explored the following input guesses of the velocity dispersions for the narrow and broad components (\sign~and \sigb): \sign = 30, 50, 80, \sigs, and \sigb=80, 150, 250, 2$\times$\sigs, 3$\times$\sigs, where \sigs~is the velocity dispersion measured from the single-component fit. 

\subsection{Identifying double-component galaxies}
We then categorized the sample of emission-line galaxies into three groups based on the presence and reliability of the broad component. We calculate the fraction, \fb, of the total line flux attributed to the broad component as:
\begin{equation} f_{\rm B} = \frac{F_{\rm H\alpha,B}}{(F_{\rm H\alpha,N}+F_{\rm H\alpha,B})}, \end{equation}
where $F_{\rm H\alpha,N}$ and $F_{\rm H\alpha,B}$ are the H$\alpha$ emission line fluxes of the narrow and broad components, respectively, obtained through the double-component fit. We identified single-component galaxies when either the $\overline\chi^2_{\text{S}}$ value was lower than the $\overline\chi^2_{\text{D}}$ value or the fraction of the broad component was less than 10\% (\fb~$<0.1$). Note that the identification of most single-component galaxies relied on the \fb~$<0.1$ criterion, as they tend to display a slightly better fit with double components, despite having insignificant flux in their second component.

Double-component galaxies were identified based on several criteria. Firstly, their $\overline\chi^2_{\text{D}}$ value had to be lower than the $\overline\chi^2_{\text{S}}$ value, indicating a better fit for the double-component model. Additionally, the fraction of the broad component had to be between 10\% to 90\% ($0.1 \le f_{\text{B}}\le0.9$), indicating a significant contribution from both components. Although the 10\% fraction cut is arbitrary, the results throughout the paper remain stable even when this fraction cut is slightly varied (e.g., $0.2 \le f_{\text{B}}\le0.8$). Furthermore, for reliable measurements, all four optical emission lines used for diagnostics (H$\beta$, [O{\small III}], H$\alpha$, and [N{\small II}]) had to exhibit A/N$>$3 for each individual component. Lastly, the velocity dispersion of the broad component ($\sigma_{\rm B}$) had to exceed the velocity dispersion of the narrow component ($\sigma_{\rm N}$) by at least its uncertainty ($\sigma_{\rm B,error}$). By applying these criteria, we confidently identified double-component galaxies, allowing us to investigate the properties and characteristics of the broad and narrow components within these systems. Galaxies not meeting the criteria for either a single- or double-component galaxy were classified as `uncertain'. Applying these criteria, the sample consists of 386 single-component galaxies, 356 double-component galaxies, and 543 uncertain galaxies. In Appendix~\ref{sec:bic}, we explore classification using the Bayesian Information Criterion (BIC); we conclude that using the BIC does not change the main results of this study. 

Uncertain galaxies show significantly lower A/N in their emission lines compared to those from single- and double-component galaxies. For instance, the median A/N ratios of the H$\alpha$ line are 332$\pm$58,  453$\pm$84, and 184$\pm$39 for single-component, double-component, and uncertain galaxies, respectively. Consequently, among the 543 uncertain galaxies, 299 exhibit A/N$<$3 for at least one of the four lines in their broad component, with 82\% of them associated with the low A/N of the H$\beta$ line. Additionally, 448 uncertain galaxies do not satisfy the condition of $\sigma_{\rm B}- \sigma_{\rm B,error} > \sigma_{\rm N}$, and 69\% of them show $\sigma_{\rm B,error}$ higher than $\sigma_{\rm B}$. Therefore, we infer that uncertain galaxies in this study do exhibit a preference for double components in describing their emission lines. However, their measured line fluxes and kinematics, especially for the broad components, lack the accuracy required for this study. For further analysis, we focussed solely on comparing single- and double-component galaxies to maintain simplicity and avoid uncertainty regarding the reality of the narrow and broad components. 

\section{Result}
\subsection{Characteristics of double-component galaxies}
\label{sec:hist}
\begin{figure*}
\centering
\includegraphics[width=\textwidth]{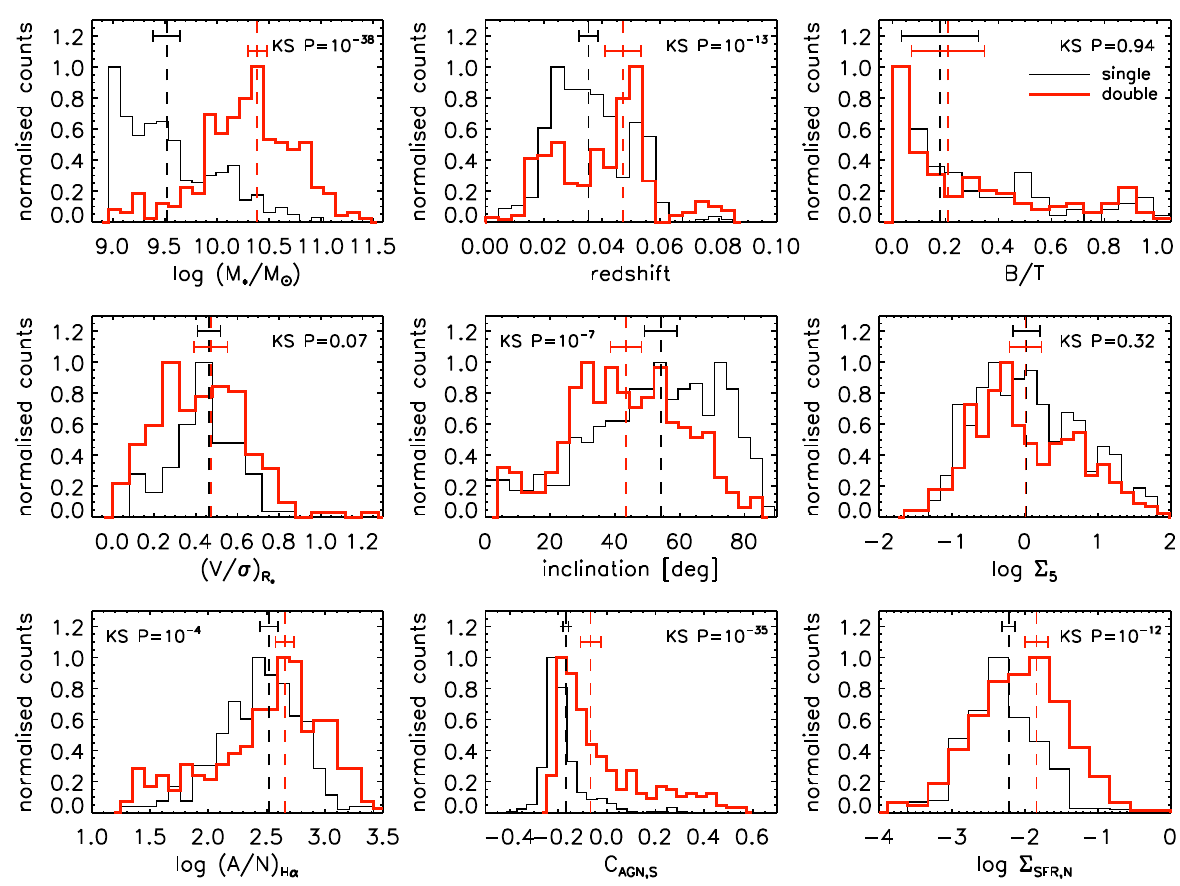}
\caption{The distributions of properties for single-component (black) and double-component (red) galaxies. Vertical dashed lines indicate the medians of the distributions, with error bars representing the 95\% confidence interval for the median. Additionally, the probability values from the Kolmogorov-Smirnov (KS) two-sample test are indicated in each panel. Single- and double-component galaxies exhibit distinct distributions in $\rm M_*$, inclination, redshift, \sfdenn, and \cagns.}
\label{hist}
\end{figure*}

We present the characteristics of single- and double-component galaxies in this section. Most of the parameters presented in this section have been obtained from the SAMI DR3 catalogue. For each parameter, we provide a concise description along with the primary reference source. The bulge-to-total (B/T) ratio has been obtained from the catalogue published by Barsanti et~al.\ (2021) and Casura et~al.\ (2022). These studies employed ProFit, a Bayesian two-dimensional galaxy profile modelling routine developed by Robotham et~al.\ (2017), to derive the B/T ratio. The stellar $\rm (V/\sigma)_{R_e}$ ratio was determined by calculating the flux-weighted mean within 1\re, using the 2-dimensional velocity and velocity dispersion maps (van de Sande et~al.\ 2017), following the approach outlined by Cappellari et~al.\ (2007). The inclination was derived using the axial ratio ($b/a$) and intrinsic flatness ($q_0$) of 0.2 following the conversion:
\begin{equation}
\cos(i) = \sqrt{\frac{(b/a)^2-q_0^2}{1-q_0^2}} ~.
\end{equation}
 The fifth-nearest neighbour surface density $\Sigma_5$ is determined using galaxies with absolute $r$-band magnitudes $\rm M_r<-18.6$ (Brough et~al.\ 2017). The \cagn\ parameter serves as a quantitative measure of the relative contribution of power sources to the ionized gas using optical emission-line diagnostics, the so-called BPT diagnostics (Baldwin, Phillips \& Terlevich 1981; Veilleux \& Osterbrock 1987). We adopted the definition of \cagn\ as described by Oh et~al.\ (2022): the shortest orthogonal departure, on a logarithmic scale, from the empirical demarcation line established by Kauffmann et~al.\ (2003) in the BPT diagram (as demonstrated in Section~\ref{sec:bpt}). We use \cagns, derived using emission line ratios measured from the single-component fit in this section, and later use that for individual components. 

The star formation rate (SFR) surface density within 1\re~(\sfden) was derived by using the dust-corrected H$\alpha$ emission flux. The derivation assumes an intrinsic Balmer decrement of $\rm H\alpha/H\beta = 2.86$ along with the Cardelli extinction law (Cardelli et al. 1989) and incorporates empirical calibrations based on Kennicutt et al. (1994) with adjustments for the Chabrier (2003) initial mass function, as detailed in Medling et~al.\ (2018). We specifically use \sfdenn\ as a proxy for star-forming activities, derived from the H$\alpha$ emission flux measured from the narrow component, to avoid potential contamination in total H$\alpha$ emission in non-star-forming galaxies. We also explored \sfden\ using the spectral energy distribution (SED) fitting method described in Ristea et~al.\ (2022), which is less susceptible to AGN contamination. Although there are some systematic changes in the range of two \sfden\, these alterations lead to only minimal changes in the results, particularly when discussing the connection between \sfden\ and gas kinematics. Therefore, although \sfdenn\ may provide upper limits for the true \sfden\ for AGN, the results with \sfdenn\ in this study are less likely artificially originated by the contamination of H$\alpha$ flux by AGN activities. We decided to use \sfden\ based on the H$\alpha$ flux because we lack SFR measurements from SED fitting for 45 galaxies in the sample, and estimating \sfden~for individual emission components is not feasible. 

The distributions of various galaxy parameters in single- and double-component galaxies are shown in Figure~\ref{hist}. The median values highlight the statistically distinct distributions observed in $\rm M_*$, redshift, inclination, \sfdenn, and \cagns\ between the two groups. We specifically examined whether the 95\% confidence intervals of the medians in the two groups overlap. Single-component galaxies are predominantly composed of low-mass galaxies, while double-component galaxies exhibit higher mass distributions. There might be a bias in identifying low-mass galaxies with double components due to the challenges in resolving their emission components, especially in blue wavelengths at the SAMI resolution limit. Single-component galaxies are more populated at lower redshifts, likely because SAMI employed multiple volume-limited samples, which introduces selection effects, leading to a higher frequency of low-mass galaxies at lower redshifts. Double-component galaxies are more frequently found in galaxies with lower inclinations, possibly because the outflow component is less detectable in edge-on galaxies. Additionally, double-component galaxies demonstrate higher values of \sfdenn\ and \cagns, suggesting a connection between the presence of a broad component and the ionization source of the emission lines. Our results confirm the distinct distributions between single- and double-component galaxies in stellar mass, inclination, and star-formation rate surface density, as reported by Zovaro et al. (2024) for the star-forming sample. They also discussed possible observational biases on multi-component emission line fits.
 
On the other hand, double-component galaxies share similarities with single-component galaxies in terms of both photometrically-defined (B/T) and kinematically-defined ($\rm(V/\sigma)_{R_e}$) galaxy types. Additionally, they exhibit similar values for local density $\Sigma_5$, pointing to internal processes as the primary physical mechanisms driving the formation of the broad component. We emphasize that double-component galaxies are not specifically associated with galaxies exhibiting high A/N ratios of H$\alpha$ emission. This result implies that the presence of double-component galaxies is not strongly influenced by, or a selection effect related to, the quality of spectra. The majority of the integrated spectra, after the velocity correction, exhibit an (A/N)$_{\rm H\alpha}$ higher than 100 for both single- and double-component galaxies. Moreover, low A/N galaxies are predominantly classified as uncertain galaxies and are subsequently excluded from the final sample. Consequently, the classification of galaxies as single- or double-component is not strongly dependent on the quality of the spectra, at least in this study.

\subsection{Emission line diagnostics}
\label{sec:bpt}
\begin{figure}
\centering
\includegraphics[width=\columnwidth]{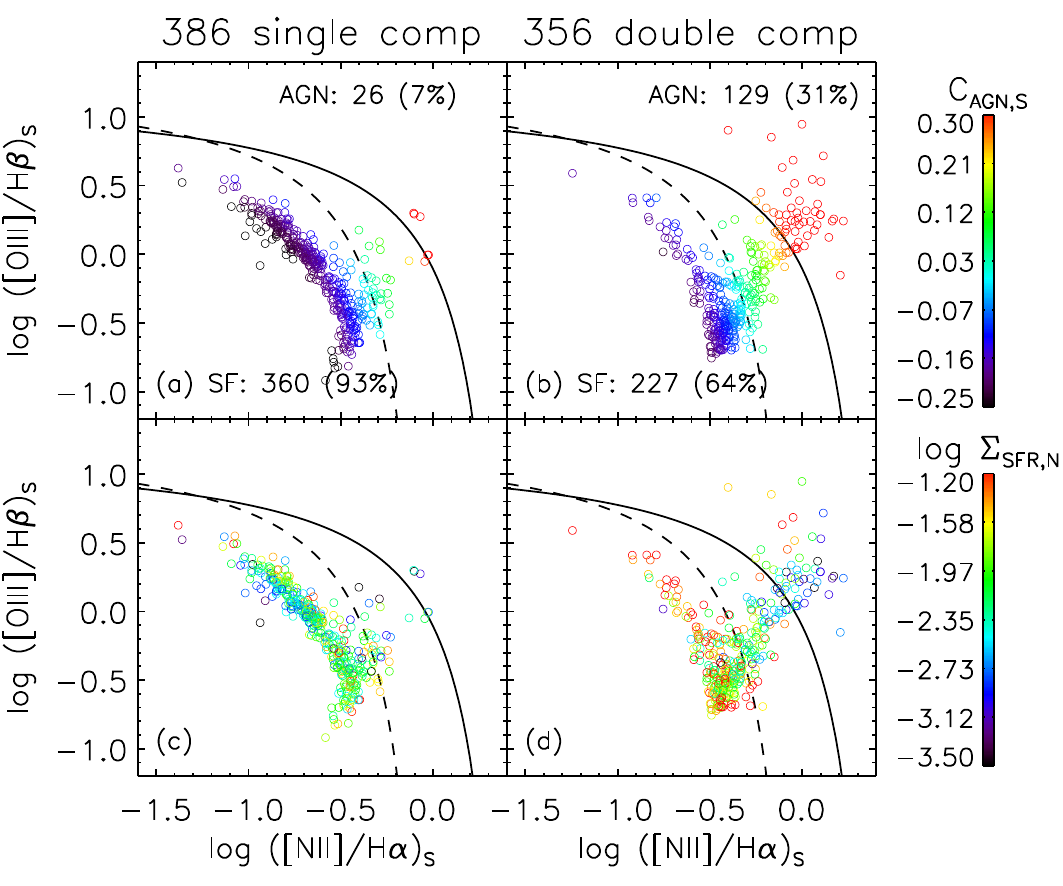}
\caption{Emission-line diagnostics for the single-component (left) and double-component (right) galaxies. Emission line fluxes in this figure are estimated from single-component fitting. The solid and dashed lines represent, respectively, the demarcation lines of Kewley et~al.\ (2001) and Kauffmann et~al.\ (2003), which distinguish between star-forming galaxies (SF) and AGN. Points are colour-coded by either \cagns\ (top) or \sfdenn\ (bottom). Double-component galaxies are more common when \cagns>0 or \sfdenn\ is high.}
\label{bpt}
\end{figure}

We employ emission-line diagnostics based on the emission-line flux ratios derived from single-component fitting (Figure~\ref{bpt}). The emission-line diagnostics diagram highlights the differences in emission line ratios between single- and double-component galaxies. The vast majority (93\%) of single-component galaxies are found among star-forming galaxies below the dashed demarcation line (\cagns<0). The remaining single-component galaxies are primarily situated within the region associated with composite galaxies, lying between the two demarcation lines established by Kewley et~al.\ (2001) and Kauffmann et~al.\ (2003). We rarely observe AGN-like emission from single-component galaxies positioned above the solid demarcation line defined by Kewley et~al.\ (2001). In contrast, double-component galaxies exhibit a more diverse distribution: 23\% and 13\% of them are associated with composite and AGN classifications, respectively. 

From another perspective, non-star-forming galaxies (\cagns>0) often exhibit a broad component. Out of the 155 galaxies with \cagns>0, a substantial majority, 129 (83\%), are classified as double-component galaxies. These double-component non-star-forming galaxies in our sample display low \sfdenn\ values, as indicated in panel~(d). The median log~\sfdenn\ values for double-component galaxies among the non-star-forming (\cagns>0) and star-forming (\cagns <0) groups are $-$2.30 and $-$1.67, respectively, with a standard deviation of 0.51\,dex. This result independently supports the notion that the mechanisms contributing to the broad component in the \cagns>0 sample are less likely associated with star-forming activities. 

On the other hand, the 590 star-forming galaxies (\cagns<0) in our sample are distributed between single-component (61\%) and double-component (39\%) galaxies. Star-forming galaxies with double components tend to exhibit higher \sfdenn\ values compared to those with a single component. Specifically, the median log~\sfdenn\ values for star-forming galaxies with a single component and those with double components are -2.21 and -1.67, respectively, with a standard deviation of 0.47\,dex. This result suggests that a broad component is more commonly observed in star-forming galaxies with higher \sfdenn\ values. For additional insights, refer to the study by Zovaro et~al.\ (2024), which also presents distinct distributions of \sfden\ in single and double-component galaxies based on a spatially resolved classification.

In conclusion, galaxies tend to exhibit a broad component when their \sfdenn\ is high or when they display AGN-like emissions (\cagns>0), suggesting multiple contributors for generating the broad component. However, as illustrated in Figure~\ref{hist}, the stellar mass emerges as the most decisive factor, having the most significant differences in the distributions between single- and double-component galaxies. In the next section, we further explore the impact of \cagns\ and \sfdenn\ on the structures and kinematics of ionized gas emission beyond stellar mass. 

\subsection{Impact of \sfden\ and \cagn\ on gas $\sigma$}
\label{sec:msig}
\begin{figure*}
\centering
\includegraphics[width=\textwidth]{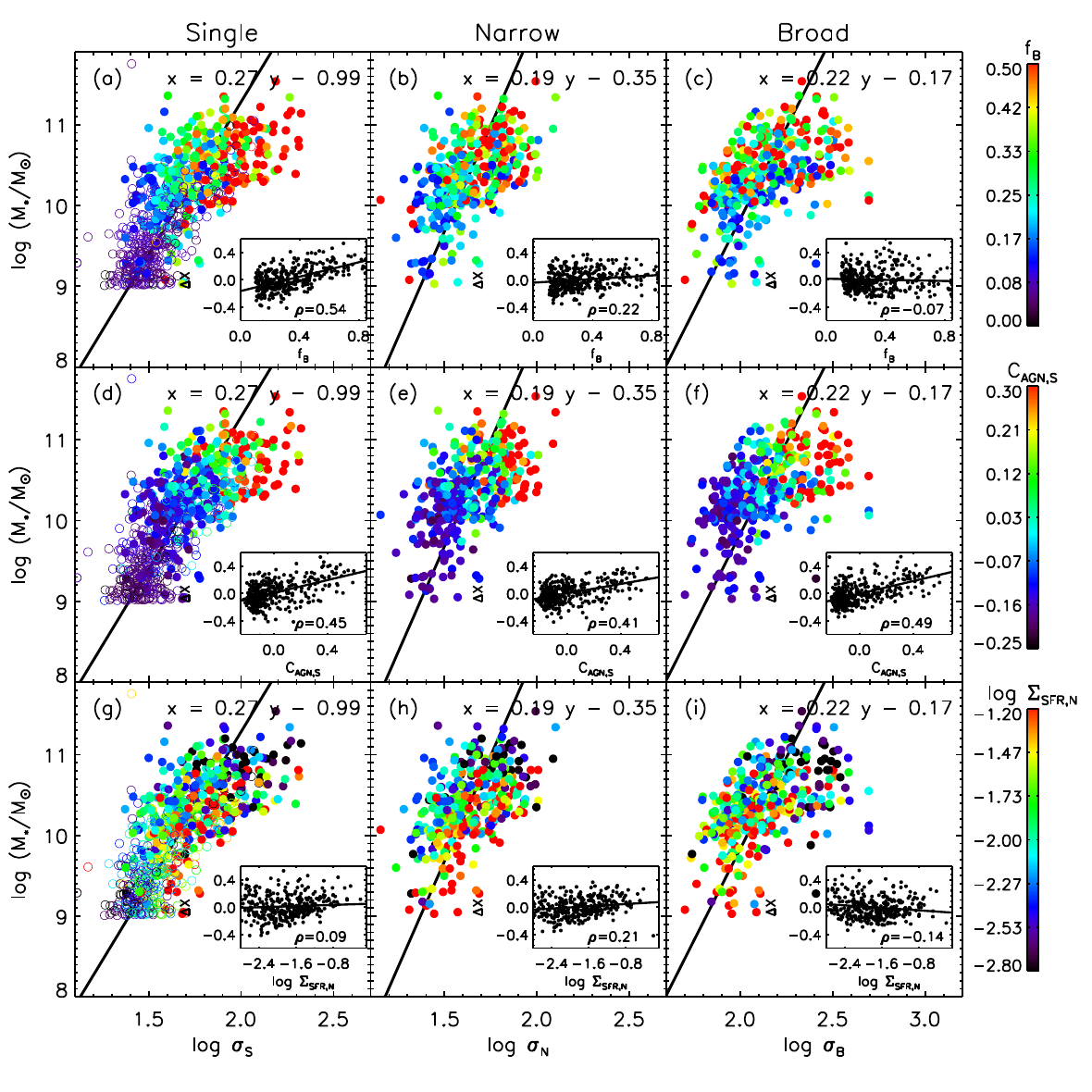}
\caption{The relations between stellar mass and gas velocity dispersion are shown for the single (left), narrow (middle), and broad (right) components; only double-component galaxies (filled circles) are considered for the narrow and broad components (middle and right columns). The data points are colour-coded according to \fb\ (a--c), \cagns~(d--f), and \sfdenn~(g--i). The best-fit line (black) is derived by fitting each velocity dispersion to the stellar mass. The inset panels show the residuals in velocity dispersion from the best fit ($\Delta$X) as a function of \fb, \cagns\ and \sfdenn, along with the Spearman coefficient ($\rho$) for the residuals. The solid line is the best fit for the residual.}
\label{msig}
\end{figure*}

In Figure~\ref{msig}, we present the relations between the stellar mass and the ionized gas velocity dispersions measured for single (\sigs), narrow (\sign), and broad component (\sigb). We display data points for both single-component galaxies (open circles) and double-component galaxies (filled circles) when examining \sigs\ in panels~(a), (d) and~(g). For panels involving either \sign\ or \sigb, we only consider double-component galaxies. We determine the best-fit line in each panel by fitting velocity dispersions to the stellar mass, aiming to analyze any deviations from the velocity dispersion predictions offered by the stellar mass. The insets in each panel provide the residual trends in velocity dispersion with respect to \fb, \cagns\ and \sfdenn. The Spearman coefficient ($\rho$) is displayed in each inset panel to quantify the significance of the residual. There are no discernible residual trends with metallicity and the other parameters presented in the previous section; thus, they were not included in the analysis.

Figure~\ref{msig}(a) shows there is a significant residual trend in \sigs\ with \fb. This result suggests that galaxies with a more pronounced presence of the broad emission-line component tend to exhibit higher \sigs\ than is predicted by their mass alone. The residual trend with \fb\ is not clearly evident in \sign\ and \sigb\ (panels b and c), implying that the influence of \fb\ on the kinematics of the narrow and broad components is relatively limited. Hence, the residual trend observed in \sigs\ with \fb\ does not originate from the individual components in isolation but rather emerges from their cumulative effect. The influence of \fb\ on \sigs\ becomes more pronounced due to the significantly higher value of \sigb\ compared to \sign. Panel~(d) introduces another substantial residual trend in \sigs\ relating to \cagns. It is noteworthy that the observed residual trend with \cagns\ extends its influence to \sign\ and \sigb\ as well (panels~e and~f). This result suggests that, unlike \fb, \cagns\ is connected to the kinematics of both the broad and narrow components. The presence of a residual trend in \sigs, \sign\ and \sigb\ when examining \cagns\ indicates that power sources have a significant role in regulating the dynamics of the entire emission-line system within galaxies. We do not observe any residual trend in gas velocity dispersions~with \sfdenn\ (panels~g--i). 

\begin{figure*}
\centering
\includegraphics[width=\textwidth]{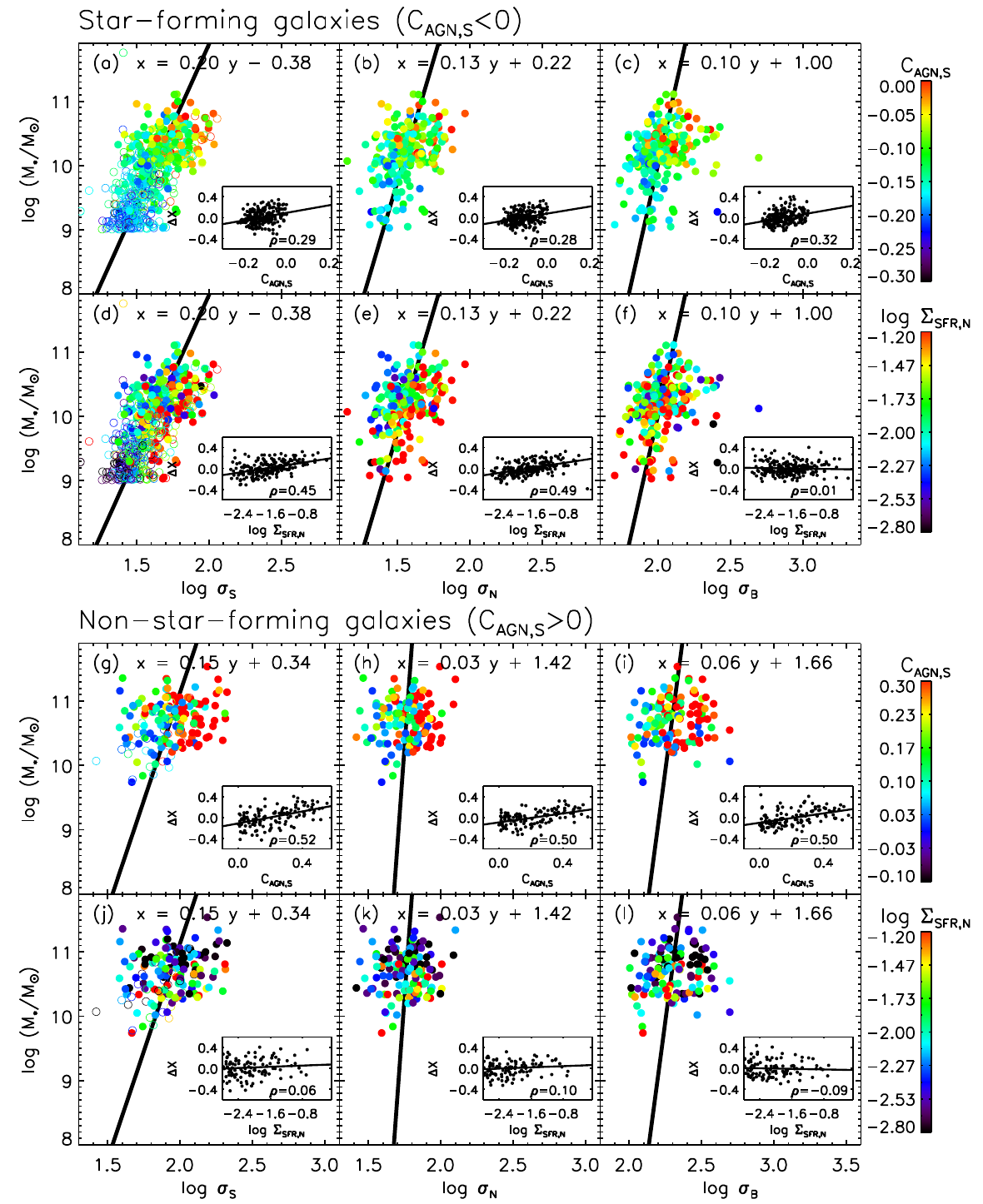}
\caption{The stellar mass and velocity dispersion relations. Details are as in Figure~\ref{msig}, but star-forming and non-star-forming galaxies are separated. Both samples display positive residual trends in gas velocity dispersions associated with \cagns\ (panels a--c; g--i). However, only star-forming galaxies show a connection between \sign\ (or \sigs) and \sfdenn\ (panels~d and~e).}
\label{msig2}
\end{figure*} 

In Figure~\ref{msig2}, we present the relationships between the stellar mass and the gas velocity dispersion within subsamples categorized by their dominant power sources. Panels (a)--(f) display star-forming galaxies with \cagns<0, while panels (g)--(l) present non-star-forming galaxies exhibiting AGN-like emissions (\cagns>0). Our results confirm a well-established observational connection between the ionization sources and stellar mass (e.g.\ Juneau et~al.\ 2011): non-star-forming galaxies are predominantly found among massive galaxies (log\,(\mstar)>10), while star-forming galaxies are distributed more evenly across the entire mass range. Substantial and persistent residual trends in gas velocity dispersions with \cagns\ are evident among non-star-forming galaxies (panels g--i). We do not notice any correlations between residual velocity dispersions and \sfdenn\ in panels (j)--(l), confirming that the kinematics of non-star-forming galaxies defined by \cagns>0 are, indeed, independent of star-forming activities. 

Even among star-forming galaxies with \cagns<0, we continue to observe positive residual trends in gas velocity dispersions associated with \cagns\ (panels a--c), though less pronounced compared to the whole sample. This attenuation may result from the limited variation in \cagns\ for star-forming galaxies. The connection between the residual in \sigs\ and \sfdenn\ only becomes evident when the sample is restricted to the star-forming galaxies (panel~d). This residual trend observed in \sigs\ appears to primarily originate from the narrow component rather than the broad component. Specifically, the narrow components exhibit a notable residual trend with \sfdenn\ (panel~e), while the broad component does not display such a trend (panel~f). However, when non-star-forming galaxies are included, the residual trend with \sfdenn\ becomes diluted, resulting in the absence of residual trends in Figure~\ref{msig}(g)--(i).

\section{Discussion}
\subsection{M$_*$--\,gas $\sigma$ relations}
\label{sec:msig}
As shown in Figures~\ref{msig} and \ref{msig2}, significant correlations between stellar mass and gas velocity dispersions are detected, particularly in star-forming galaxies. Therefore, in the previous section, we examined the impact of \cagn\ and \sfden\ as proxies for the power of star formation and AGN-driven feedback, excluding the influence of stellar mass on gas velocity dispersions. In this section, we discuss the reasons behind the significant correlation between stellar mass and ionized gas velocity dispersions. 

The stellar mass is expected to be proportional to the global potential of a galaxy. In this context, one may anticipate a correlation between the stellar mass and gas kinematics, given that ionized gas is also a part of a galaxy influenced by the gravitational potential. Recent studies employing IFS have indeed reported positive correlations between stellar mass and the gas velocity dispersion (or emission line width) (Cortese et~al.\ 2014; Barat et~al.\ 2019, 2020; Rodr{\'\i}guez del Pino et~al.\ 2019; Swinbank et~al.\ 2019; Varidel et~al.\ 2020; Oh et~al. 2022). However, the connection between gas kinematics and the global potential is expected to be less prominent, owing to their sensitivity to local physics, including turbulences in the interstellar medium (e.g.\ Federrath \& Klessen 2012), star-formation feedback (e.g.\ Lehnert et~al.\ 2009; Green et~al.\ 2010), AGN feedback (e.g.\ Guillard et~al.\ 2015); gas accretion (e.g.\ Klessen \& Hennebelle 2010), gravitational instability (e.g.\ Krumholz \& Burkert 2010); and thermal contamination of H\,\textsc{ii} region (e.g.\ Krumholz et~al.\ 2018). 

SAMI data, in general, are limited by seeing effects, with beam smearing playing a significant role in the observed correlation between gas velocity dispersions and stellar mass. Beam smearing spatially smooths line-of-sight radial velocity distributions from the IFS data, resulting in artificially elevated velocity dispersions and decreased velocity gradients. Varidel et~al.\ (2020) estimated intrinsic gas kinematics for 383 star-forming galaxies correcting for the impact of beam smearing using a 3D forward-modelling technique. Figure 3 of Varidel et~al.\ (2020) highlights that corrected gas velocity dispersions are significantly lower than the observed velocity dispersion. Nevertheless, a positive correlation still exists between the corrected gas velocity dispersions and stellar mass. However, their M$_*$--\,gas $\sigma$ correlation is less pronounced, with a Spearman coefficient of 0.31, compared to the correlation shown in this study (see Figure~\ref{msig2}(a) with a Spearman coefficient of 0.65) for star-forming galaxies. We therefore suspect that beam smearing artificially amplifies the M$_*$--\,gas $\sigma$ correlation, by convolving rotational velocity (a proxy for mass) into the velocity dispersion.

The magnitude of beam smearing increases primarily when either the size of the seeing relative to the galaxy size or the velocity gradient is high. The former condition implies a greater impact on small galaxies, given the small variations in seeing in the SAMI data. We estimated the local gas velocity gradient $\nabla V$ at each x, y coordinate using the line-of-sight velocity values of neighbouring spaxels (V), following the approach outlined in Varidel et~al. (2016):
\begin{equation} \nabla V(x,y) \equiv \frac{1}{2 W_{\rm pix}}\sqrt{\begin{aligned} &[V_{\rm }(x+1,y)-V_{\rm }(x-1,y)]^2 \\
     & + [V_{\rm }(x,y+1)-V_{\rm }(x,y-1)]^2
    \end{aligned}
    }~, \end{equation}
where $W_{\rm pix}$ is the size of the SAMI spaxel (0.5 arcsecs). We find a strong positive correlation between stellar mass and $\nabla V$, with Spearman coefficient of 0.61, thereby confirming the influence of beam smearing on the M$_*$--gas $\sigma$ relation presented in this study (see also Zhou et~al.\ 2017). 

We investigated whether the influence of beam smearing extends to the roles of \sfden\ and \cagn, as illustrated in Figure~\ref{msig2}. In Appendix~\ref{sec:vgrad}, we present a result similar to Figure~\ref{msig2}, replacing stellar mass with $\nabla V$ and considering $\nabla V$ as a proxy for the impact of beam smearing. We observed that, even after accounting for the $\nabla V$--\sign\ relation in Figure~\ref{vgrad}(e), there is still a robust residual correlation between \sfdenn\ and \sign. Furthermore, we detected residual correlations between gas $\sigma$ and \cagn\ beyond the $\nabla V$--\sign\ relation (Figure~\ref{vgrad}g-i). Additionally, there were no significant changes in the results throughout this study when replacing stellar mass with $\nabla V$. Therefore, we conclude that star-formation and AGN-driven outflows associated with \sfden\ and \cagn\ are contributing to the inflation of gas velocity dispersions even after accounting for the impact of beam smearing. See also Lehnert et~al.\ (2013) reporting a positive connection between \sfden\ and gas $\sigma$ after correcting for the impact of beam smearing. 

Our analysis involving $\nabla V$ is not intended to suggest that the M$_*$--gas $\sigma$ relation originates entirely from beam smearing. Although we use $\nabla V$ as an indicator for the magnitude of beam smearing, it is a conservative approach because $\nabla V$ is also a physically motivated quantity and is not determined solely by observational limitations. For instance, galaxies with compact H$\alpha$ emissions or outflows may exhibit steep velocity gradients. Furthermore, previous studies have confirmed a positive correlation between stellar mass and gas $\sigma$, even after correcting for beam smearing. Therefore, we interpret the M$_*$--gas $\sigma$ relation as being partially driven by the dependence of gas kinematics on the global potential, but the relation has been artificially strengthened by beam smearing. In the following sections discussing the narrow and broad component kinematics, we continue to use stellar mass to account for the impact of both global potential and beam smearing on gas kinematics. 

\subsection{Narrow component kinematics}
\begin{figure*}
\centering
\includegraphics[width=\textwidth]{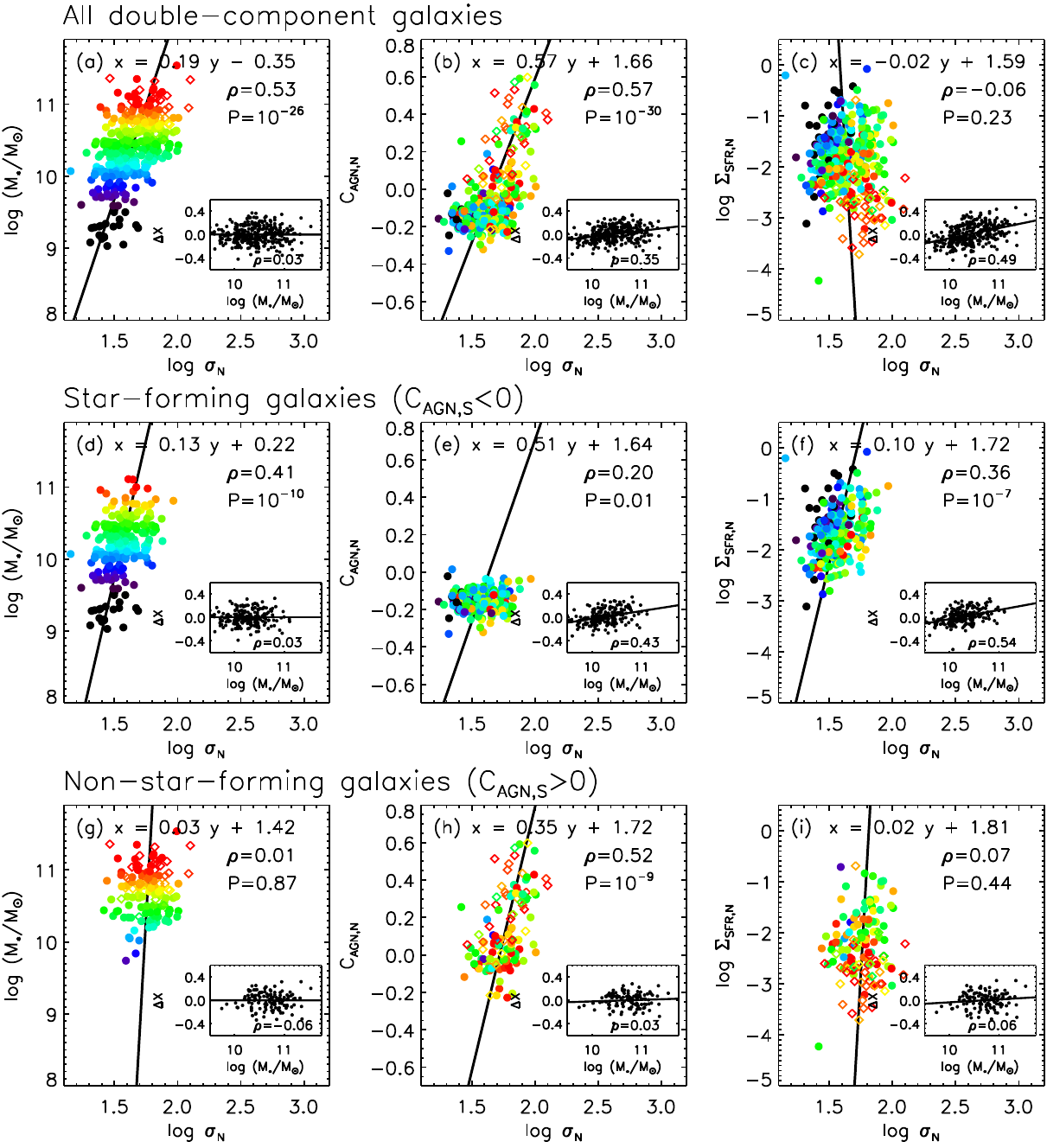}
\caption{The dependences of \sign\ on stellar mass, \cagnn\ and \sfdenn. We present results for all double-component galaxies in panels (a)--(c), star-forming galaxies in panels (d)--(f), and non-star-forming galaxies in panels (g)--(i). The data points are colour-coded according to stellar mass (see panels a, d, and g), and the insets display the residuals in log~\sign\ from the best fit as a function of stellar mass. Retired galaxies with $\rm EW(H\alpha)_S$ below 3\,\AA are represented by diamond symbols. We provide Spearman coefficient $\rho$ and its significance (P) for each panel. Other details are as in Figure~\ref{msig}. In star-forming galaxies, \sign\ independently correlates with stellar mass and \sfdenn. On the other hand, only \cagnn\ contributes to \sign\ for non-star-forming galaxies.}
\label{sign}
\end{figure*} 

Figures~\ref{msig} and \ref{msig2} suggest that, along with stellar mass, both \cagn\ and \sfden\ contribute to gas kinematics in certain cases. In this section, we further examine the correlations between the velocity dispersion of the narrow component and the key parameters (Figure~\ref{sign}). We utilise \cagnn\ and \sfdenn\ measured from the line fluxes of the narrow component, but the results remain consistent even when we employ \cagns\ and \sfdens\ instead. In Figure~\ref{sign}, panels~(a) and~(b) indicate that both stellar mass and \cagnn\ display strong positive correlations with \sign, as evidenced by their Spearman coefficients (0.54 and 0.57, respectively). In contrast, panel~(c) reveals no discernible correlation between \sfdenn\ and \sign. However, when considering all sample galaxies, these findings do not provide a clear explanation for the influence of key parameters on gas kinematics, particularly with regard to \sfdenn. To address this, we further divided the sample into star-forming galaxies (panels d-f) and non-star-forming galaxies (panels h-j).

Star-forming galaxies still exhibit a strong positive correlation between stellar mass and \sign\ (Figure~\ref{sign}d). The insets in panels~(d) and~(e) suggest that stellar mass describes \sign\ better than \cagnn\ for star-forming galaxies. Panel~(f) clearly shows a positive correlation between \sign\ and \sfdenn, which was not detected in panel~(c) when considering all sample galaxies. The inset in panel~(f) also reveals a residual trend with respect to stellar mass. By integrating the results from panels~(d)--(f) in Figure~\ref{sign} with panel~(e) in Figure~\ref{msig2}, we conclude that both stellar mass and \sfdenn\ independently contribute to the narrow-component kinematics of star-forming galaxies. The results suggest an association between the narrow component in star-forming galaxies and the gravitational potential, establishing a correlation between stellar mass and the velocity dispersion of ionized gas. However, it is important to consider that the connection with the gravitational potential may be significantly lower than indicated by the M$_*$--\sign\ relation, given the impact of beam smearing discussed in the previous section. Additionally, the influence of \sfdenn\ indicates that star-formation-driven outflows perturb the motion of ionized gas within the narrow component. This result supports previous studies reporting positive correlations between the gas $\sigma$ and \sfden\ (e.g.\ Lehnert et~al.\ 2009, 2013; Green et~al.\ 2010; Varidel et~al.\ 2020).

Among our sample, non-star-forming galaxies are predominantly massive galaxies (log(\mstar)>10), as shown in Figure~\ref{sign}(g). For these galaxies, narrow-component kinematics are primarily dependent on \cagnn\ (panel~h). The inset in panel~(h) confirms that there is no residual trend with respect to stellar mass beyond the correlation between \cagnn\ and \sign, indicating that stellar mass does not effectively describe \sign\ in non-star-forming galaxies, probably due to their limited mass range. The dependence of \sign\ on \cagnn\ for non-star-forming galaxies may provide evidence for the impact of AGN-induced turbulence (e.g.\, Guillard et~al.\ 2015; Carniani et~al.\ 2017; Wittor \& Gaspari 2020). Additionally, non-star-forming galaxies do not exhibit any dependency of \sign\ on \sfdenn\ (panel~i). The influence of \sfdenn\ is only observed in the narrow-component kinematics of star-forming galaxies, while non-star-forming galaxies obscure the connection between star formation and gas kinematics. 

We further examine the narrow component of non-star-forming galaxies by considering the equivalent widths of the H$\alpha$ emission line measured from the single-component fit ($\rm EW(H\alpha)_S$), which is often used to identify retired galaxies that have ceased star formation (e.g.\ Cid Fernandes et~al.\ 2010, 2011). We find that all star-forming galaxies show $\rm EW(H\alpha)_S$ values above 5\,\AA. On the other hand, 59\% of non-star-forming galaxies with $\rm EW(H\alpha)_S$>3\,\AA\ are expected to be associated with AGN, and their narrow component may be associated with the `narrow line regions' (NLR) in AGN. The remaining 41\% of non-star-forming galaxies have $\rm EW(H\alpha)_S$ values below 3\,\AA, indicating that they can be classified as retired galaxies. Consequently, the narrow component in the retired galaxies does not appear to be associated with star formation or young stars; instead, it is powered by old stars. 

Figure~\ref{sign}(g)-(i) highlights that retired galaxies typically exhibit high stellar masses (log(\mstar) $\gtrsim 10.5$), low \sfdenn\ values (log\sfdenn\ $\lesssim-2.5$) and \cagnn\ values broadly distributed above 0. The retired galaxies show a positive correlation between \cagnn\ and \sign\ (Figure~\ref{sign}h), similar to AGN with $\rm EW(H\alpha)_S > 3$\,\AA, suggesting that old stellar populations also contribute to the narrow component kinematics. This contribution is possibly due to hot old post-AGB (post-asymptotic giant branch) stars (Singh et al. 2013). However, it is less likely to be from extraplanar diffuse ionized gas (Belfiore 2016), as the majority of retired galaxies in this study are massive early-type galaxies. However, it is important to note that selecting retired galaxies using the $\rm EW(H\alpha)_S$ measured within 1 \re\ may introduce a bias. AGN emissions are often highly centrally concentrated, and using integrated spectra with a large aperture (e.g., 1 \re) can obscure the AGN contribution, leading to an overidentification of retired galaxies.

\subsection{Broad component kinematics}
\begin{figure*}
\centering
\includegraphics[width=\textwidth]{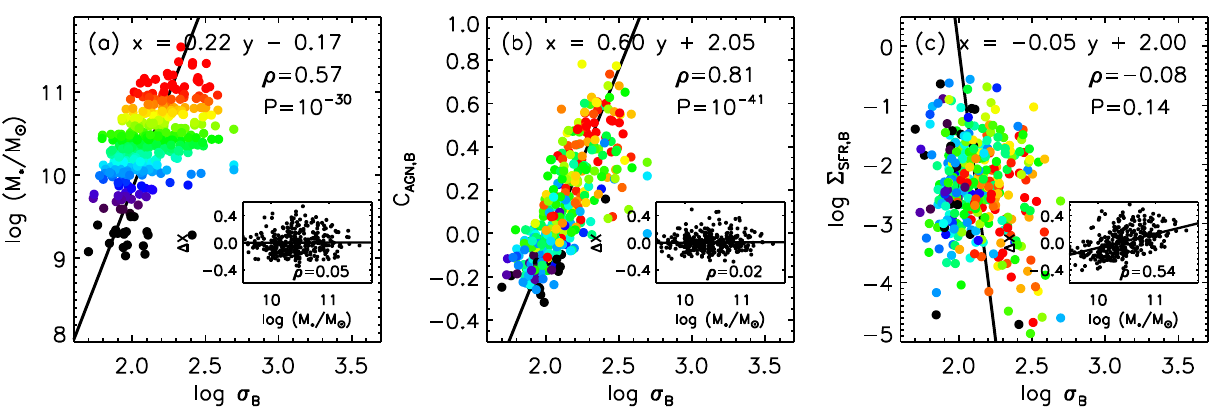}
\caption{The dependence of \sigb\ on stellar mass, \cagnb\ and \sfdenb, for double-component galaxies. We do not present results for subgroups because no additional correlations were identified to distinguish between star-forming and non-star-forming galaxies. Details are as in Figure~\ref{sign}. Panel (b) emphasizes that \cagnb\ determines \sigb, and there is no remaining correlation with stellar mass once accounting for the correlation between \sigb\ and \cagnb.}
\label{sigb}
\end{figure*} 

In this section, we further investigate the correlation between the velocity dispersion of the broad component and key parameters. In Figure~\ref{sigb}, panel~(b) clearly demonstrates a strong correlation between \sigb\ and \cagnb, underlining \cagnb\ as the primary driver of broad-component kinematics. Importantly, the inset of panel~(b) reveals no residual correlation between $\Delta$(log\,\sigb) and stellar mass. Thus, the observed correlation between stellar mass and \sigb\ in panel~(a) is secondary, stemming from the strong link between stellar mass and \cagnb~(for further observational evidence, see Juneau et al. 2011 and Vitale et al. 2013). Furthermore, simulations indicate a positive correlation between instantaneous AGN power and stellar mass, as illustrated in Figure 7 of Beckmann et al. (2017). This correlation may stem from the positive relationship between stellar and black hole masses and/or high accretion rates (e.g.\ Silk \& Norman, 2009). The result provides compelling evidence that broad components are disconnected from the gravitational potential but are significantly influenced by non-gravitational feedback processes, such as galactic outflows. Note that even after excluding retired galaxies with $\rm EW(H\alpha)<3$, the results remain consistent, with a Spearman coefficient of 0.79 for the correlation between \sigb\ and \cagnb.

There is no positive correlation between \sigb\ and \sfdenb, indicating that the kinematics of the broad component remain largely unaffected by the level of star formation (panel~c). In contrast to narrow-component kinematics, differentiating between star-forming and non-star-forming galaxies when analyzing \sfdenb\ does not reveal hidden relationships. For non-star-forming galaxies, we do not anticipate any correlations between \sigb\ and \sfdenb\ since their broad component is not associated with star formation processes. However, even in star-forming galaxies, no significant correlation between \sigb\ and \sfdenb\ is visible. We also conducted the same analysis using \sfdenn\ instead of \sfdenb, considering its significance as an indicator of star formation activity, as broad components may not appear to be closely tied to star-forming processes, particularly in galaxies influenced by AGN or dominated by old stars. We obtained consistent results when using \sfdenn, not finding any correlations between \sigb\ and \sfdenn\ even for star-forming galaxies ($\rho=-0.08$). 

It is important to emphasize that the absence of the direct correlation between \sigb\ and \sfdenn\ does not necessarily indicate that the broad components in star-forming galaxies are entirely disconnected from star-forming activities. In fact, we do observe broad components in star-forming galaxies, particularly when their \sfdenn\ is elevated. Furthermore, the emission line ratios of the broad components closely resemble those of star-forming (i.e.\ \cagnb<0) or composite galaxies in the BPT diagnostic diagram. Therefore, we can rule out AGN and old stars as the primary sources of the broad component in star-forming galaxies. One plausible scenario is that active star formation triggers gas outflows, leading to the formation of a broad emission-line component. However, even more intense star-forming activities may only increase the magnitude of the outflows (or \fb), but may not increase the velocity of the outflow, which therefore does not exert a significant additional influence on the gas kinematics (Hopkins et~al.\ 2018). 

\section{Summary and Conclusion}
We investigated the influence of star-formation and AGN-driven outflows on the kinematics of ionized gas in 1285 galaxies from the SAMI galaxy survey, employing both broad and narrow emission-line components. We generated velocity-field-corrected aperture spectra within 1\re, as detailed in Section~\ref{sec:spectra}. We conducted a simultaneous fitting of strong emission lines using double Gaussian (i.e.\ broad and narrow) components with the {\sc{lzifu}} software, which is specifically tailored for SAMI data, as described in Section~\ref{sec:fit}. We categorized the sample into single- and double-component galaxies based on the presence and reliability of the broad component, considering the $\overline\chi^2$ of the fit, the flux contribution of the broad component (\fb), and the amplitude-to-noise (A/N) ratios for H$\alpha$, H$\beta$, [N\,{\small II}], and [O\,{\small III}] emission lines. Applying these criteria yields 386 single-component galaxies and 356 double-component galaxies. 

Double-component galaxies, exhibiting both broad and narrow emission-line components, tend to be massive, with high star-formation rate surface density \sfdenn\ or \cagns\ values, which quantifies the contribution from AGN emission (see Figure~\ref{hist}). However, there might be an observational bias (e.g.\ low A/N and spectral resolution) that results in the preferential selection of double-component galaxies among massive galaxies, as discussed in Section~\ref{sec:hist}. Our analysis of the BPT diagnostic diagram, along with \sfdenn\ and \cagns, further confirms that the prevalence of broad components is more pronounced in galaxies displaying stronger star-forming activities or AGN-like emissions (refer to Figure~\ref{bpt}; see also Avery et~al. 2021).

A significant $M_*$--gas $\sigma$ relation as shown in Figure~\ref{msig} may encompass both influences from the dependence of gas kinematics on the global potential and beam smearing (Section~\ref{sec:msig}). Substantial $\sigma$ residuals are observed in the $M_*$--gas $\sigma$ relation, varying with the broad component fraction (\fb), \cagns, and \sfdenn~(Section~\ref{sec:msig} and Figure~\ref{msig}). The prominence of the broad component significantly contributes to the elevation of the gas velocity dispersion in single-component measurements (\sigs). However, \fb\ does not drive an increase in the velocity dispersion of the narrow (\sign) and broad component (\sigb), suggesting that the prominence of the broad component does not impact the kinematics of individual components. In contrast, \cagns\ positively correlates with the residual $\sigma$ in both narrow and broad components, indicating that galaxies with AGN-like emissions exhibit elevated velocity dispersions in both components. 

The impact of star formation feedback, quantified by \sfden, becomes apparent when the sample is limited to star-forming galaxies (\cagns<0; see Figure~\ref{msig2}). In star-forming galaxies, the narrow component displays a significant $\sigma$ residual that positively correlates with \sfden~(see also, e.g.\ Lehnert et~al.\ 2013; Varidel et~al.\ 2020), while the broad component does not show a correlation between the $\sigma$ residual and \sfden. In contrast, non-star-forming galaxies (\cagns>0) do not display any correlations between the residual $\sigma$ and \sfden, confirming their independence from star-forming activities.

The velocity dispersion of the narrow component depends on stellar mass, \cagn\ and \sfden, but this dependence varies with their primary emission sources. In star-forming galaxies, both stellar mass and \sfden\ independently contribute to \sign\ (see Figure~\ref{msig2}e and Figure~\ref{sign}f). The result suggests that the narrow component kinematics may be partially explained by the gravitational potential, and more vigorous star-forming activity exerts an additional influence, further increasing \sign. 

On the other hand, for non-star-forming galaxies neither stellar mass nor \sfden\ determine \sign; instead, it primarily correlates with \cagn. Notably, most non-star-forming galaxies have \cagnn, measured from the narrow component, greater than 0, confirming that the narrow component in the non-star-forming galaxies is not associated with star-forming activities. We infer that the narrow component in 41\% of the non-star-forming galaxies with EW(H$\alpha$)<3\,\AA\ is linked to old stellar populations. For the remaining non-star-forming galaxies with EW(H$\alpha$)>3\,\AA, the narrow component is likely associated with the narrow line region of AGN. 

The broad emission-line component is predominantly driven by feedback processes, significantly contributing to the inflation of the gas velocity dispersion. \sigb\ is mainly governed by \cagnb\ (Figure~\ref{sigb}). Importantly, there is no residual trend with stellar mass in the \cagnb--\sigb\ relation (see the inset of Figure~\ref{sigb}b), suggesting that the $M_*$--\sigb\ relation is secondary, originating from the connection between the stellar mass and \cagn. The absence of a residual correlation between stellar mass and \sigb\ indicates that the broad component is generated independently of the gravitational potential through non-gravitational feedback-driven processes, such as galactic outflows. Especially for non-star-forming galaxies (\cagns>0), we suspect AGN feedback is the main driver for galactic outflow, effectively inflating the gas velocity dispersion. 

On the other hand, \sigb\ does not exhibit correlations with \sfden\ even in star-forming galaxies (Figure~\ref{sigb}c), suggesting that active star formation does not directly elevate the velocity dispersion of the broad component. However, it is important to note that the result does not imply that the broad components in star-forming galaxies are entirely unrelated to star-forming activities. Broad components in star-forming galaxies still present similar emission line ratios to typical star-forming galaxies in the BPT emission line diagnostics, ruling out a connection to AGN or old stars. Moreover, powerful star formation promotes the presence of the broad component, as evidenced by a higher frequency of double-component galaxies when \sfden\ is high (Figures~\ref{hist} and \ref{bpt}). A plausible scenario involves that active star formation triggers gas outflows, resulting in the formation of a broad emission-line component. However, even more intense star-forming activities, which may increase the magnitude of outflows, do not significantly impact the velocity of the outflows and, consequently, the gas velocity dispersion.

In this study, we examined the impact of feedback processes from AGN and star-forming activities on the global gas velocity dispersion integrated within 1\,\re. AGN mainly contribute to the broad emission-line component, likely through gas outflows, leading to an increase in the gas velocity dispersion. More potent AGN, identified by high \cagnb, have a more pronounced effect on inflating the gas velocity dispersion, especially in the broad component. Active star formation also elevates the global gas velocity dispersion, achieved by increasing the velocity dispersion of the narrow component and generating the broad component. However, the influence of star-forming activities on the kinematics of ionized gas is considerably less effective compared to that of AGN.

\section*{Acknowledgements}
This research and the SAMI Galaxy Survey were supported by the Australian Research Council Centre of Excellence for All Sky Astrophysics in 3 Dimensions (ASTRO~3D; ARC project number CE170100013), the Australian Research Council Centre of Excellence for All-sky Astrophysics (CAASTRO; ARC project number CE110001020), and by participating institutions. The SAMI Galaxy Survey is based on observations made at the Anglo-Australian Telescope. The SAMI instrument was developed jointly by the University of Sydney and the Australian Astronomical Observatory. The SAMI input catalogue is based on data taken from the Sloan Digital Sky Survey, the GAMA survey, and the VST ATLAS survey. The SAMI Galaxy Survey website is sami-survey.org. This research used the {\sc pPXF} method and software by Cappellari \& Emsellem (2004), as upgraded in Cappellari (2017). SO acknowledges support from the Korean National Research Foundation (NRF) (RS-2023-00214057), as well as ongoing support from DL. S.K.Y. acknowledges support from the Korean NRF (2020R1A2C3003769, 2022R1A6A1A03053472). AR acknowledges that this research was carried out while the author was in receipt of a Scholarship for International Research Fees (SIRF) and an International Living Allowance Scholarship (Ad Hoc Postgraduate Scholarship) at The University of Western Australia. We thank Wonjae Yee for exploring emission line diagnostics for individual components.

\section*{Data Availability}

The data used in this study are available from Astronomical Optics' Data Central service at https://datacentral.org.au/ as part of the SAMI Galaxy Survey Data Release 3.

\appendix
\section{Identifying double-component galaxies using Bayesian Information Criterion}
\label{sec:bic}
\begin{figure*}
\centering
\includegraphics[width=\textwidth]{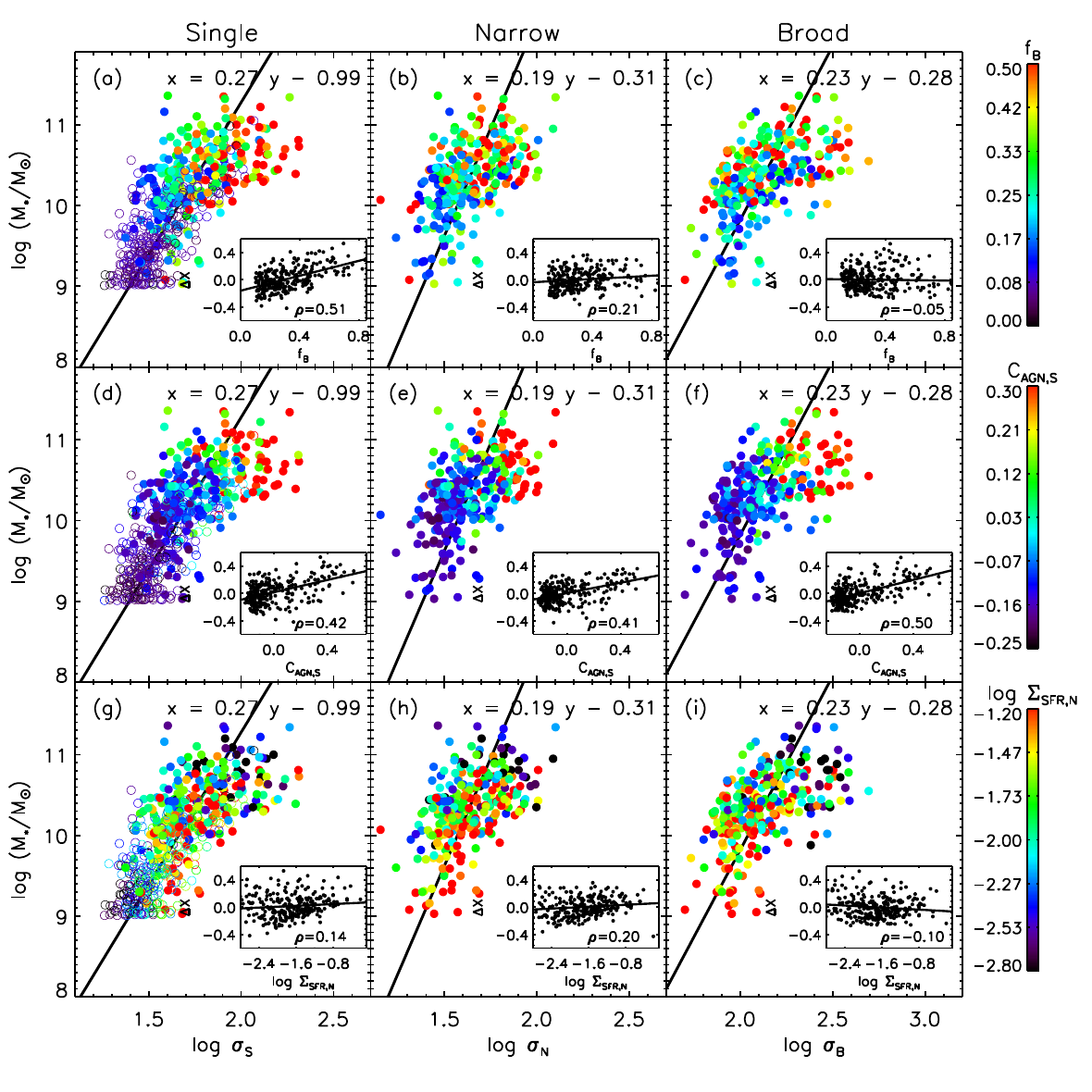}
\caption{The relations between stellar mass and gas velocity dispersions. Details are as in Figure~\ref{msig}, but the sample galaxies are classified into single- and double-component galaxies based on the BIC. No qualitative changes are observed when compared to Figure~\ref{msig}.}
\label{msig_bic}
\end{figure*} 

We explored the classification of single- and double-component galaxies using the Bayesian Information Criterion (BIC) as an alternative to relying on $\overline\chi^2$. The BIC evaluates both the goodness of fit and the complexity of the model, appropriately penalizing more complex models, such as the double-component fit considered in this study. We calculated the BIC for the single- and double-component fits, denoted as BIC(single) and BIC(double). Single-component galaxies were classified when BIC(single) $-$ BIC(double) $<$ 10 and \fb $<0.1$. Double-component galaxies were classified based on the criterion, BIC(single) $-$ BIC(double) $>$ 10, while also ensuring that \fb $>0.1$ and A/N for all four emission lines was above 3. The remaining sample, not included in the two groups, was classified as uncertain galaxies. Using the BIC classification, the sample consisted of 323 single-component galaxies, 268 double-component galaxies, and 694 uncertain galaxies. The number of double-component galaxies decreased by 24\% when applying the BIC classification due to the penalty imposed on the double-component fit. Therefore, it provides a conservative method for identifying double-component galaxies.

In Figure~\ref{msig_bic}, we present the relationships between stellar mass and gas velocity dispersions, similar to Figure~\ref{msig}, with classifications based on the BIC as described above. When comparing classifications based on $\overline\chi^2$ and BIC, no substantial differences are observed between Figures~\ref{msig} and~\ref{msig_bic}. Reproducing all the analyses based on the BIC classifications revealed only minimal changes in the results. The classification based on $\overline\chi^2$ was ultimately chosen for inclusion in the main text because the BIC classification significantly reduced the number of double-component galaxies without causing any qualitative changes in the main results.

\section{Impact of \sfden\ and \cagn\ on gas $\sigma$ beyond beam smearing}
\begin{figure*}
\centering
\includegraphics[width=\textwidth]{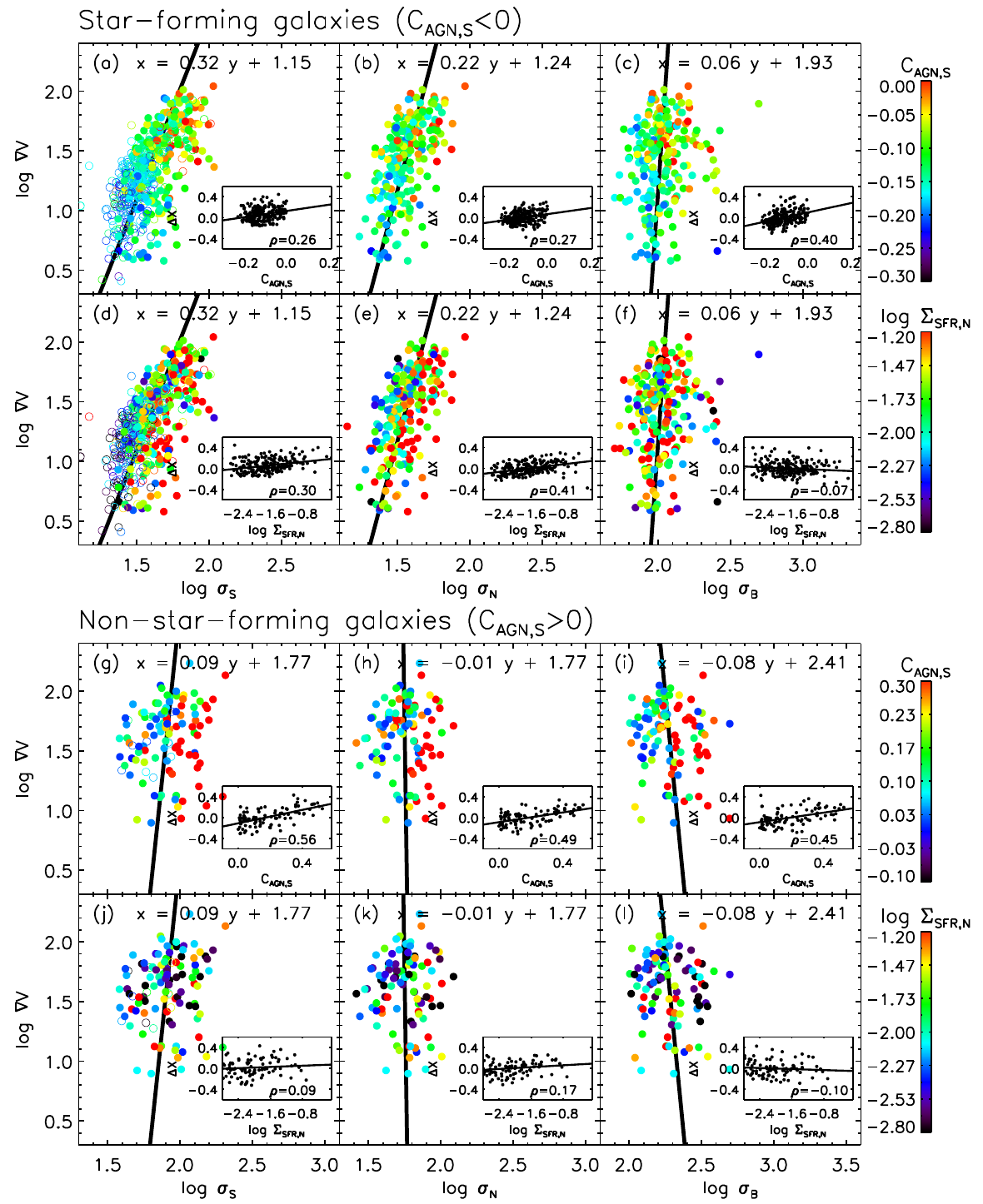}
\caption{The relations between the gas velocity gradient ($\nabla V$) and gas velocity dispersions for star-forming and non-star-forming galaxies. Other details remain consistent with those in Figure~\ref{msig2}.}
\label{vgrad}
\end{figure*}

In Figure~\ref{vgrad}, we present the $\nabla V$--$\sigma$ relation to illustrate the influence of \cagns\ and \sfdenn\ on the gas velocity dispersions, extending beyond the effects of beam smearing. The residuals in velocity dispersions, according to \cagns\ and \sfdenn, shown in the $M_*$--$\sigma$ relation in Figure~\ref{msig2}, are also displayed in the $\nabla V$--$\sigma$ relation. Refer to Section~\ref{sec:msig} for a comprehensive discussion.

\label{sec:vgrad}

\label{lastpage}
\end{document}